\documentclass[sigconf,natbib=true,screen=true, anonymous=false, review=false]{acmart}

\PassOptionsToPackage{dvipsnames}{xcolor}
\usepackage{tablefootnote}
\usepackage{dsfont}
\usepackage{acronym}
\usepackage{amsmath}
\usepackage[linesnumbered,lined,ruled,commentsnumbered]{algorithm2e}
\usepackage{bbm}
\usepackage{bm}
\usepackage{booktabs}
\usepackage[skip=0pt]{caption}
\usepackage[T1]{fontenc}
\usepackage{graphicx}
\usepackage{hyperref}
\usepackage{listings}
\usepackage{multirow}
\usepackage{natbib}
\usepackage{subcaption}
\usepackage{tabularx}
\usepackage[dvipsnames]{xcolor}
\usepackage{enumerate}
\usepackage[inline]{enumitem}
\usepackage{numprint}
\usepackage[normalem]{ulem}
\usepackage{mleftright}
\usepackage{xcolor}
\usepackage{physics}

\acrodef{GNN}{graph neural network}
\acrodef{IR}{information retrieval}
\acrodef{LTR}{learning to rank}
\acrodef{MAB}{multi-armed bandit}
\acrodef{MC}{Markov chain}
\acrodef{NBR}{next basket recommendation}
\acrodef{PIF}{personal item frequency}
\acrodef{RNN}{recurrent neural network}
\acrodef{TREx}{two-step repetition-exploration}
\acrodef{TREx-Rep}{TREx with repetition module only}
\acrodef{SOTA}{state-of-the-art}

\newcommand{\OurMethod}{TREx}

\acrodef{AWRF}{Attention-Weighted Rank Fairness}
\acrodef{EUR}{Exposed Utility Ratio}
\acrodef{RUR}{Realized Utility Ratio}

\acrodef{IAA}{Inequality of Amortized Attention}
\acrodef{EEL}{Expected Exposure Loss }
\acrodef{EER}{Expected Exposure Relevance}
\acrodef{EED}{Expected Exposure Disparity}
\acrodef{logDP}{Demographic Parity}

\acrodef{logEUR}{Exposed Utility Ratio}

\acrodef{logRUR}{Realized Utility Ratio}

\makeatletter
\g@addto@macro\normalsize{%
  \abovedisplayskip 2pt plus1pt %
  \belowdisplayskip 2pt plus1pt
  \abovedisplayshortskip  0pt plus1pt%
  \belowdisplayshortskip  0pt plus1pt%
}
\makeatother

\newcommand{\header}[1]{\vspace{0.8mm}\noindent\textbf{#1.}}

\author{Ming Li}
\authornote{Both authors contributed equally to the paper.}
\orcid{0000-0001-7430-4961}

\author{Yuanna Liu}
\authornotemark[1]
\orcid{0000-0002-9868-6578}
\affiliation{%
        \institution{University of Amsterdam}
        \city{Amsterdam}
        \country{The Netherlands}
}
\email{m.li@uva.nl, y.liu8@uva.nl}

\author{Sami Jullien
}
\orcid{0000-0003-4507-6335}
\affiliation{%
        \institution{AIRLab, University of Amsterdam}
        \city{Amsterdam}
        \country{The Netherlands}
}
\email{s.jullien@uva.nl}

\author{Mozhdeh Ariannezhad}
\orcid{0000-0002-1113-8094} 
\affiliation{%
        \institution{Booking.com}
        \city{Amsterdam}
        \country{The Netherlands}
}
\email{mozhdeh.ariannezhad@booking.com}
\authornote{Work done while the author was with AIRLab, University of Amsterdam.}

\author{Andrew Yates
}
\orcid{0000-0002-5970-880X} 
\affiliation{%
        \institution{University of Amsterdam}
        \city{Amsterdam}
        \country{The Netherlands}
}
\email{a.c.yates@uva.nl}

\author{
Mohammad Aliannejadi
}
\orcid{0000-0002-9447-4172} 
\affiliation{%
        \institution{University of Amsterdam}
        \city{Amsterdam}
        \country{The Netherlands}
}
\email{m.aliannejadi@uva.nl}

\author{Maarten de Rijke}
\orcid{0000-0002-1086-0202} 
\affiliation{%
        \institution{University of Amsterdam}
        \city{Amsterdam}
        \country{The Netherlands}
}
\email{m.derijke@uva.nl}

\renewcommand{\shortauthors}{Li et al.}

\copyrightyear{2024} 
\acmYear{2024} 
\setcopyright{rightsretained} 
\acmConference[SIGIR '24]{Proceedings of the 47th International ACM SIGIR Conference on Research and Development in Information Retrieval}{July 14--18, 2024}{Washington, DC, USA}
\acmBooktitle{Proceedings of the 47th International ACM SIGIR Conference on Research and Development in Information Retrieval (SIGIR '24), July 14--18, 2024, Washington, DC, USA}
\acmDOI{10.1145/3626772.3657835}
\acmISBN{979-8-4007-0431-4/24/07}

\ccsdesc[500]{Information systems~Recommender systems}
\ccsdesc[300]{Information systems~Retrieval models and ranking}

\keywords{Next basket recommendation; Repetition and exploration; Evaluation}

\begin{document}
\title[TREx]{Are We Really Achieving Better Beyond-Accuracy Performance\\ in Next Basket Recommendation?}

\begin{abstract}

\Ac{NBR} is a special type of sequential recommendation that is increasingly receiving attention.
So far, most \acs{NBR} studies have focused on optimizing the accuracy of the recommendation, whereas optimizing for beyond-accuracy metrics, e.g., item fairness and diversity remains largely unexplored.
Recent studies into \ac{NBR} have found a substantial performance difference between recommending repeat items and explore items. 
Repeat items contribute most of the users' perceived accuracy compared with explore items. 

Informed by these findings, we identify a potential ``short-cut'' to optimize for beyond-accuracy metrics while maintaining high accuracy.
To leverage and verify the existence of such short-cuts, we propose a plug-and-play \acfi{TREx} framework that treats repeat items and explores items separately, where we design a simple yet highly effective repetition module to ensure high accuracy, while two exploration modules target optimizing \textit{only} beyond-accuracy metrics. 

Experiments are performed on two widely-used datasets w.r.t.\ a range of beyond-accuracy metrics, viz.\ five fairness metrics and three diversity metrics.
Our experimental results show that:
\begin{enumerate*}[label=(\roman*)]
    \item we can achieve \acl{SOTA} performance w.r.t.\ accuracy via the designed repetition module in \ac{TREx}; and 
    \item the simple \ac{TREx} framework achieves ``better'' beyond-accuracy performance than existing sophisticated methods.
\end{enumerate*}
Prima facie, this appears to be good news: we can achieve high accuracy \emph{and} improved beyond-accuracy metrics at the same time.
However, we argue that the real-world value of our algorithmic solution, \ac{TREx}, is likely to be limited and reflect on the reasonableness of the evaluation setup.
We end up challenging existing evaluation paradigms, particularly in the context of beyond-accuracy metrics, and provide insights for researchers to navigate potential pitfalls and determine reasonable metrics to consider when optimizing for accuracy and beyond-accuracy metrics.
\end{abstract}

\maketitle

\acresetall

\section{Introduction}
Recommender systems have become an essential instrument for connecting people to the content, services, and products they need. 
In e-commerce, more and more consumers purchase food and household products online instead of visiting physical retail stores~\citep{kumar-2017-measuring}. 
The COVID-19 pandemic has only accelerated this shift~\citep{oecd-2020-e-commerce}. 
In this scenario, consumers usually purchase a \emph{set} of items at the same time, a so-called \emph{basket}. 
\Acfi{NBR} is a type of sequential recommendation that caters to this scenario: baskets are the target of recommendation and historical sequential data consists of users' interactions with baskets. 
\Ac{NBR} has increasingly been attracting attention in recent years~\citep{ariannezhad-2023-complex}. 
Many methods, based on different machine learning techniques, have been proposed for accurate recommendations, e.g., \ac{MC}-based methods~\citep{fpmc, hrm}, frequency and nearest neighbor-based methods~\cite{tifuknn, recency}, \acs{RNN}-based methods~\cite{dream, beacon, sets2sets, clea}, and self-attention methods~\citep{dnntsp, sun2020timestamp2, chen2021modelingcat}.

\header{Repetition vs.\ exploration in \ac{NBR}}
Recently, \citet{nbr-rep-expl} have assessed the performance of \acl{SOTA} \ac{NBR} in terms of repeat and explore items: items that a user has interacted with before and items that they have never interacted with before, respectively.
The authors distinguish between the task of repetition recommendation (recommending \emph{repeat items}) and the task of exploration recommendation (recommending \emph{explore items}).
Repetition and exploration recommendations have different levels of difficulty, where recommending items that are regularly present in a user’s baskets is shown to be a far easier task~\cite{nbr-rep-expl}. 
Building on these findings, repetition-only~\citep{buycycle, recanet} and exploration-only~\citep{li2023masked} methods have been proposed to optimize the accuracy of next basket recommendation. 

\header{Accuracy and beyond-accuracy metrics}
Even though accuracy naturally serves as the most important objective of recommendations, it is widely recognized that it should not be the sole focus. Beyond-accuracy metrics such as item fairness~\citep{10.1145/3331184.3331380, 10.1145/3404835.3462882, 10.1145/3437963.3441824, 10.1145/3477495.3532007} and diversity~\citep{chen2021multi, zhang2008avoiding, zhao2023fairness} also play crucial roles in evaluating recommendation services.
Such beyond-accuracy metrics have gained increasing attention and have been optimized in a range of recommendation scenarios~\cite{yin2023understanding, zhao2023fairness}.
In the \ac{NBR} scenario, however, beyond-accuracy metrics have been far less studied than accuracy-based metrics.
In this paper, we help to address this knowledge gap. 
Following the paradigm of multiple-objective recommender systems~\citep{jannach-2022-multi-objective}, it is widely recognized that there is a trade-off between accuracy and beyond-accuracy metrics. E.g., diversity goals are reckoned to stand in contrast with accuracy. 
Put differently, a method achieving a better beyond-accuracy performance while maintaining the same level of accuracy performance is considered to be a  success~\citep{yin2023understanding, zhao2023fairness}. 
And how can we achieve a reasonable balance between accuracy and beyond-accuracy metrics in \ac{NBR}?

\header{Potential ``short-cuts'' to balancing accuracy and beyond-accuracy metrics} Besides the imbalance between repetition and exploration~\citep{nbr-rep-expl, li-2023-repetition, li-2023-repetition-offline, li-2023-will}, \citeauthor{nbr-rep-expl} also found that repeat items contribute most of the accuracy, whereas the explore items in the recommended basket contribute very little to the user's perceived utility. 
As Table~\ref{tab:diff-rep-expl} summarizes, there are essential differences between the repetition and exploration tasks, which explain the substantial performance differences between the two tasks.

Inspired by these findings, we hypothesize that there may be a ``short-cut'' strategy to optimize for both accuracy and beyond-accuracy metrics, which contains two aspects:
\begin{enumerate*}[label=(\roman*)]
    \item \emph{accuracy}: Predict repeat items to achieve good accuracy: predicting repeat items is much easier than predicting explore items~\citep{nbr-rep-expl}, and 
    \item \emph{beyond-accuracy}: Use explore items to improve beyond-accuracy metrics:  it is very difficult to recommend quality explore items. Thus, exchange the low accuracy that is typically achieved on such items for beyond-accuracy metrics, i.e., trade accuracy for diversity and item fairness.
\end{enumerate*}
We call this \ac{NBR} strategy a short-cut strategy because it avoids making the fundamental trade-off between accuracy and beyond-accuracy metrics.

\begin{table}[t]
\centering
\newcommand{\tabincell}[2]{\begin{tabular}{@{}#1@{}}#2\end{tabular}}
  \caption{Comparison of the repetition and exploration tasks in \ac{NBR}.}
  \setlength{\tabcolsep}{2pt}
  \label{tab:diff-rep-expl}
  \begin{tabular}{lll}
    \toprule
     \bf Aspect & \bf Repetition & \bf Exploration \\
    \midrule
    \em Task difficulty & Easy & Difficult \\
    \midrule
    \em Number of items & Dozens & Thousands \\
    \midrule
    \em Item interactions & Previous & None \\
    \cmidrule{1-3}
    \em Users' interest & With feedback  & Without feedback  \\
    \cmidrule{1-3}
    \em Task type & Re-consume & Infer new \\
    \bottomrule
  \end{tabular}
\end{table}

\header{TREx framework}
To operationalize our short-cut idea, and check whether the ``short-cut'' strategy can be made to work, we propose the \acfi{TREx} framework. 
\ac{TREx} decouples the prediction of repeat items and explore items.
Specifically, \acs{TREx} uses separate models for predicting
\begin{enumerate*}
    \item[(a)] repeat items, and
    \item[(b)] explore items, 
\end{enumerate*}
and then combines the outcomes of the two prediction models to generate the next basket.
In contrast, existing \acs{NBR} methods usually output the scores/probabilities of all items and then select the top-$k$ items to fill up a basket to be recommended, ignoring the differences between repeat and explore items.

For \ac{TREx}'s repeat item prediction,  we propose a simple yet effective probability-based method, which considers the item characteristics and users' repurchase frequency. For exploration recommendations, we design two strategies that cater to the different beyond-accuracy metrics.
The flexibility of \acs{TREx} allows us to design suitable models for repetition and exploration, with the possibility of controlling the proportions of repetition and exploration to investigate the relations between accuracy and various beyond-accuracy metrics.

\header{Findings and reflections}
We consider two types of widely-used beyond-accuracy metrics, i.e., diversity and item fairness. 
Specifically, we investigate five fairness metrics (i.e., logEUR, logRUR, EEL, EED, and logDP)~\citep{liu-2024-measuring, raj2022measuring} and three diversity metrics (i.e., ILD, Entropy, and DS)~\citep{yin2023understanding}. 
To provide an overall understanding of these metrics, we group them according to different levels of connection with accuracy as follows: 
\begin{enumerate*}[label=(\roman*)]
    \item Strong connection: logRUR,
    \item Weak connection: logEUR, EEL, EED 
    \item No connection: logDP, ILD, Entropy, DS.
\end{enumerate*}
Briefly, the strong connection between logRUR and accuracy stems from the fact that logRUR uses ground truth relevance to discount the exposure, making sure that only correctly predicted items contribute to effective exposure. The connection between logEUR, EEL, and accuracy is weak because they just ensure the exposure distribution across groups of recommended results is close to the group exposure distribution of ground truth, without considering whether the exposure is contributed by correctly predicted items. Since the position weighting model of EED considers ground truth, EED shows a weak connection. There is no connection between accuracy and logDP, ILD, Entropy, and DS because their exposure distributions across groups are designed to reflect a specific distribution. The strength of the connection between a beyond-accuracy metric and accuracy determines whether there is a short-cut towards optimizing both accuracy and the beyond-accuracy metric. 

We perform experiments on two brick-and-mortar retailers' \acs{NBR} datasets, considering six NBR baselines and eight metrics. The experimental results show that:
\begin{enumerate*}
    \item State-of-the-art accuracy can be achieved by only recommending repeat items via the proposed simple yet effective repetition model.
    \item Leveraging the ``short-cut'' using \acs{TREx} achieves ``better'' beyond accuracy performance w.r.t.\ seven out of eight beyond-accuracy metrics.
    \item In terms of the item fairness metric having a strong connection with the accuracy (i.e., logRUR), it is more difficult to achieve better beyond-accuracy metrics via the proposed strategy.
\end{enumerate*}

\header{Stepping back}
Instead of blindly claiming \acs{TREx} with the designed modules as a \acl{SOTA} method for optimizing both accuracy and various beyond-accuracy metrics, we reflect and challenge our evaluation paradigm in the definition of success in this setting. 
The core question is:
\begin{quote}
    \emph{Are we really achieving better beyond-accuracy performance in \acl{NBR}?}
\end{quote}
Two perspectives offer different ways forward for researchers and practitioners to address this question:
\begin{enumerate}[leftmargin=*,nosep]
    \item If we are willing to sacrifice the accuracy of the exploration, then superior beyond-accuracy performance can be achieved by leveraging the ``short-cut'' strategy via \acs{TREx}, which is straightforward and efficient. This ``short-cut'' strategy must be considered before developing more sophisticated and elaborate approaches. 
    \item Conversely, if we believe it is unreasonable to sacrifice the accuracy of exploration~\citep{williams-2014-emotions}, the existence of the ``short-cut'' strategy reveals flaws in our current evaluation paradigm to demonstrate an NBR method's superiority. 
    A fine-grained analysis (i.e., distinguishing between repetition and exploration) needs to be performed to check whether ``better'' beyond-accuracy is achieved by triggering the ``short-cut'' strategy, which would hurt the exploration accuracy after all.
\end{enumerate}

\header{Our contributions}
The main contributions of the paper are:
\begin{itemize}[leftmargin=*]
    \item We identify a ``short-cut'' strategy (i.e., sacrificing accuracy for exploration and using explore items to optimize for beyond-accuracy metrics), which could achieve ``better'' beyond-accuracy metrics without degrading accuracy.
    \item We propose a simple repetition recommendation model considering item features and users' repurchase frequency, which can achieve the \acl{SOTA} NBR accuracy by only recommending repeat items.
    \item We propose \ac{TREx}, a flexible two-step repetition-exploration framework for \ac{NBR}, which allows us to control the trade-off between accuracy and beyond-accuracy metrics w.r.t. the recommended baskets.
    \item We conduct experiments on two datasets w.r.t. eight beyond-accuracy metrics, and find that leveraging ``short-cuts'' via \acs{TREx} can achieve better performance on a wide range of metrics. We also find that the stronger the connection with accuracy, the more challenging it becomes to utilize a ``short-cut'' strategy to enhance a beyond-accuracy metric.
    \item We reflect on, and challenge, existing evaluation paradigms, and find that a fine-grained level analysis can provide a complementary view of a method's performance.
\end{itemize}

\section{Related Work}
We summarize related research on \acl{NBR} and beyond-accuracy metrics.

\header{Next basket recommendation}
The \ac{NBR} problem has been studied for many years. 
Factorizing personalized Markov chains (FPMC) \citep{fpmc} leverages matrix factorization and Markov chains to model users' general interest and basket transition relations. 
HRM \citep{hrm} applies aggregation operations to learn a hierarchical representation of baskets. 
RNNs have been adapted to the \ac{NBR} task to learn long-term trends by modeling the whole basket sequence. E.g., Dream \citep{dream} uses max/avg pooling to encode baskets. Sets2Sets~\citep{sets2sets} adapts an attention mechanism and adds frequency information to improve performance. 
Some methods~\citep{beacon, intnet} consider the underlying item relations to get a better representation. 
\citet{dnntsp} argue that item-item relations between baskets are important, and leverage GNNs to use these relations.
Some methods~\citep{bai2018attribute, wang2019timestamp1, sun2020timestamp2, nbrdiversity} exploit auxiliary information, including product categories, amounts, prices, and explicit timestamps. 
TIFUKNN~\citep{tifuknn} and UP-CF@r~\citep{recency}, frequency-neighbor-based methods, model temporal patterns, and then combine these with neighbor information or user-wise collaborative filtering. 
\citet{nbr-rep-expl} provide several metrics to evaluate repetition and exploration performance in the \acs{NBR} task and find that the repetition task is easier than the exploration task. Inspired by this analysis, repetition-only~\citep{recanet, buycycle} and exploration-only~\cite{li2023masked} models were proposed for \acl{NBR}.
Existing \acs{NBR} work mainly focuses on optimizing accuracy whereas this paper extends to various beyond-accuracy metrics for \acs{NBR}.

\header{Beyond-accuracy metrics}
In addition to accuracy, there are various beyond-accuracy metrics (i.e., diversity, fairness, novelty, serendipity, coverage) we need to consider when making recommendations~\citep{10.1145/3331184.3331380}. 
Diversity is a crucial factor in meeting the diverse demands of users~\citep{zhang2008avoiding, quadrana2018sequence, chen2020improving, wang2019modeling}. 
Recently, empirical and revisitation studies~\citep{ludewig2018evaluation, yin2023understanding} have been conducted to explore the trade-off between accuracy and diversity.
The concepts of fairness and item exposure have emerged as crucial considerations since items and producers play pivotal roles within a recommender system and its ecosystem. 
Related metrics measure whether items receive a fair share of exposure according to different definitions of fairness.
Current research on fairness primarily focuses on individual or group fairness, either from the customer's perspective, adopting a user-centered approach~\citep{bobadilla2020deepfair}, or from the provider's viewpoint, adopting an item-centered approach~\citep{10.1145/3366424.3380048, 10.1145/3397271.3401100}, or a two-sided approach~\citep{10.1145/3404835.3462882, 10.1145/3477495.3532007, 10.1145/3477495.3531959}. 
Recently, \citet{liu-2024-measuring} evaluated item fairness on existing NBR methods to investigate the robustness of different fairness metrics.
Unlike the work listed above, this paper is not limited to optimizing a specific type of metric. 
It examines the possibility of leveraging a ``short-cut'' strategy to seemingly optimize various beyond-accuracy metrics and provides insights w.r.t.\ evaluation paradigms when extending \acs{NBR} optimization and evaluation to these beyond-accuracy metrics.

\begin{table}[t]
	\centering
	\newcommand{\tabincell}[2]{\begin{tabular}{@{}#1@{}}#2\end{tabular}}
	\caption{Notation used in the paper; fairness related notation is adapted from~\citep{raj2022measuring, liu-2024-measuring}.}
	\setlength{\tabcolsep}{2pt}
	\label{tab:notations}
	\begin{tabular}{ll}
		\toprule
		Symbol & Description \\
		\midrule
		$u\in U$ & Users \\
		$i\in I$ & Items \\
		$S_{u}$ & Sequence of historical baskets for $u$ \\
		$B_{u}^{t}$ & $t$-th basket in  $S_{u}$, a set of items $i \in I$ \\
  		$I_{u,t}^{rep}$ &Set of repeat items for $u$ up to timestamp $t$\\
		$I_{u,t}^{expl}$ &Set of explore items for $u$ up to timestamp $t$ \\
		$T_{u}$ & Ground-truth basket for $u$ that we aim to predict\\
		$T_{u}^\mathit{rep}$ & Set of \emph{repeat items} in the ground truth basket $T_{u}$ for $u$\\
		$T_{u}^\mathit{expl}$ & Set of \emph{explore items} in the ground truth basket $T_{u}$ for $u$\\
		$P_{u}$ & Predicted basket for $u$  \\
		$P_{u}^\mathit{rep}$ & Set of \emph{repeat items} in the predicted basket $P_{u}$ for $u$\\
		$P_{u}^\mathit{expl}$ & Set of \emph{explore items} in the predicted basket $P_{u}$ for $u$\\
$G(P)$  & Group alignment matrix for items in $P$ \\
$G^{+}$ & Popular group \\
$G^{-}$ & Unpopular group \\
$\vb{a}_P$ & Exposure vector for items in $P$ \\
$\mathbf{\epsilon}_P$ & The exposure of groups in $P$ $(G(P)^T \vb{a}_P)$\\

		\bottomrule
	\end{tabular}
\end{table}

\vspace*{-2mm}
\section{Task Formulation and Definitions}

We describe the next basket recommendation problem and formalize the notions of repetition and exploration.
Our notation is summarized in Table~\ref{tab:notations}.

\header{Next basket recommendation}
Given a set of users $U = \{u_1$, $u_2$, \ldots, $u_n\}$ and items $I = \{i_1 ,i_2, \ldots, i_m\}$, $S_u = \{B_u^1, B_u^2, \ldots, B_u^t\}$ represents the  historical interaction sequence for $u$, where $B_u^t$ is the user's basket at the time step $t$. $B_u^t$ consists of a set of items $i\in{I}$, and the goal of the \emph{next basket recommendation} task is to predict $P_{u} = B_u^{t+1}$, the following basket of items that the user would probably like, based on the user's past interactions $S_u$, i.e.,
\begin{equation}
P_{u} = \hat{B}_u^{t+1} = f(S_u),
\end{equation}
where $f$ is our basket generation algorithm. 
We assume that the user's attention and screen space is limited; hence, like previous studies~\citep{nbr-rep-expl, liu-2024-measuring}, we recommend fixed-size baskets of sizes 10 or 20.

\header{Repetition and exploration}
We assume that the set of items is fixed. 
Although this might not be the case in real-world settings, modeling the addition and deletion of items in the set of items is out of the scope of this paper. 
With this assumption in mind,  the addition of every new basket to the users' history,  may translate into fewer items left to explore. 
To differentiate between the items coming from the exploration and repeat consumption behavior, for a user $u$ and timestamp $t$, a set of items $I_{u,t}^\mathit{rep} \subset I$ are considered to be  the ``repeat items.'' The set of explore items $I_{u,t}^\mathit{expl}$ is simply its complement within the overall item set $I$.
We define $I_{u,t}^\mathit{rep}$ as:
\begin{equation}
I_{u,t}^\mathit{rep} = I_{u,t-1}^{rep} \cup B_u^t.
\end{equation}
This also means that $I_{u,1}^\mathit{rep} \subset\cdots \subset I_{u,t-1}^\mathit{rep} \subset I_{u,t}^\mathit{rep}$. Conversely, we have $I_{u,t}^\mathit{expl} \subset I_{u,t-1}^\mathit{expl} \subset \cdots \subset I_{u,1}^\mathit{expl}$.

The task of predicting the next basket for a user $u$ is equivalent to predicting which items from $I_{u,t}^{rep}$ and $I_{u,t}^{expl}$ will appear in $B_u^{t+1}$.  One way to solve this problem is to decouple it into two subtasks: the repetition subtask that aims to predict which items from $I_{u,t}^{rep}$ to recommend, and the exploration task that recommends items from $I_{u,t}^{expl}$. 
Table~\ref{tab:diff-rep-expl} shows the different characteristics w.r.t. the repetition and exploration tasks.

\begin{table*}[t]
\caption{Summary of fairness and diversity metrics; fairness metrics are adapted from \citep{raj2022measuring}. $\uparrow$ indicates that higher values are better; $\downarrow$ indicates that lower values are better; $\circ$ means that the closer the value is to 0, the better the performance.} 
\label{metric_table}
\centering
\begin{tabular}{lllcc}
\toprule
Category   & Metrics  & Goal  & Better& Accuracy connection  \\  
\midrule
\multirow{3}{*}{\begin{tabular}{l}Equal\\ opportunity\end{tabular}} 
                  & logRUR & Click-through
rate proportional to relevance  &  $\circ$ & Strong\\
&logEUR  & Exposure proportional to relevance & $\circ$ & Weak\\

                  & EEL&Exposure matches ideal (from relevance)  & $\downarrow$ & Weak\\  
\midrule
\multirow{2}{*}{\begin{tabular}{l}Statistical\\ parity\end{tabular}} & EED &  Exposure well-distributed  &$\downarrow$ & Weak \\
                  & logDP &Exposure equal across groups  &  $\circ$ & None \\

\midrule
\multirow{3}{*}{\begin{tabular}{l}Diversity\end{tabular}} & ILD & Average distance between categories for each pair of items in the list & $\uparrow$ & None \\
                  & Entropy & Entropy of item category distribution in the list & $\uparrow$ & None\\
                  & DS & Number of categories divided by the number of items in the list & $\uparrow$ & None \\

\bottomrule

\end{tabular}

\end{table*}

\section{Evaluation metrics}
\label{sec: metric}
Next, we describe the accuracy and beyond-accuracy metrics (i.e., fairness and diversity) considered in the paper.\footnote{Due to space limitations, we only provide brief introductions of each metric; more detailed information (e.g., function, responsibility, etc.) can be found in the original papers and relevant survey papers~\citep{zhao2023fairness, raj2022measuring, liu-2024-measuring}.}

\header{Accuracy}
In terms of accuracy, we use three metrics that are widely used for the \acs{NBR} task: $Recall@k$, $NDCG@k$, and $PHR@k$. $Recall$ measures the ability to find all items that the user will purchase in the next basket; $\textit{NDCG}$ is a ranking metric that also considers the order of the items; $\textit{PHR}$ is a user level measurement which represents the ratio of users whose recommended basket contains the item in the ground-truth.

\header{Fairness} 
Assume $\pi(P\mid u)$ is a user-dependent distribution and $\rho(u)$ is a distribution over users; overall, the recommended item rankings among all users follow the following distribution: $\rho(u)\pi(P\mid u)$. 
$\epsilon_P = G(P)^\mathrm{T} \vb{a}_P$ is the group exposure within a recommended basket.\footnote{The formula to compute the exposure vector $\vb{a}_P$ using different position weighting models can be found in~\citep{raj2022measuring, liu-2024-measuring}.} Its expected value $\epsilon_\pi = E_{\pi\rho}[\epsilon_P]$ is the group exposure among all the recommended baskets.
Following~\citep{raj2022measuring, liu-2024-measuring}, we select a set of well-known fairness metrics and cover two types of fairness considerations as follows:\footnote{Item fairness metric Inequity of Amortized Attention~\citep{biega2018equity} is not used in this paper since some baselines do not have predicted relevance for items.}

\vspace*{-1mm}
\subsubsection*{(1) Equal opportunity} 
Promote equal treatment based on merit or utility, regardless of group membership~\citep{raj2022measuring, liu-2024-measuring}.
\begin{enumerate*}[label=(\roman*)]
\item \emph{\ac{EUR}}~\citep{singh2018fairness} quantifies the deviation from the objective that the exposure of each group is proportional to its utility $Y\left(G\right)$.

\item \emph{\ac{RUR}}~\citep{singh2018fairness} models actual user engagement, the click-through rates for the groups $\Gamma\left(G\right)$ are proportional to their utility.

\item \emph{\ac{EEL}}~\citep{diaz2020evaluating} is the distance between expected exposure and target exposure $\mathbf{\epsilon}^\ast$, which is the exposure under the ideal policy. 
\end{enumerate*}

\vspace*{-1mm}
\subsubsection*{(2) Statistical parity} 
Ensure comparable exposure among groups.
\begin{enumerate*}[label=(\roman*)]
\item \emph{\ac{EED}}~\cite{diaz2020evaluating} measures the inequality in exposure distribution across groups.
\item \emph{Demographic Parity (DP)}~\cite{singh2018fairness} measures the ratio of average exposure given to the two groups.
Following \cite{raj2022measuring}, we reformulate DP as $\text{logDP}$ to tackle the issue of empty-group scenarios and improve interpretability. 
\ac{logEUR} and \ac{logRUR} are defined in a similar manner.
\end{enumerate*}

\header{Diversity} Following~\citep{yin2023understanding}, we consider the following widely-used diversity metrics, which satisfy users' diversified demands.
\begin{enumerate*}[label=(\roman*)]
\item \emph{Intra-List Distance (ILD)}~\citep{chen2020improving, cen2020controllable} measures the average distance between every pair of items in the recommendation list ($P_u$), where $d_{ij}$ is the Euclidean distance between the respective embeddings of categories

\item \emph{Entropy}~\citep{zheng2021dgcn, wang2019modeling} quantifies the dispersion of item category distribution in the recommendation list $P_u$; a higher degree of dispersion in the category distribution corresponds to increased diversity.

\item \emph{Diversity Score (DS)} \cite{liang2021enhancing} is calculated as the number of interacted/recommended categories divided by the number of interacted/recommended items.
\end{enumerate*}
As shown in Table~\ref{metric_table}, we can group beyond-accuracy metrics according to their connection with accuracy.

\section{A Two-Step Repetition-Exploration Framework}

Given the differences depicted in Table~\ref{tab:diff-rep-expl}, we propose a \acfi{TREx} framework for \ac{NBR}. \ac{TREx} assembles recommendations from a repetition and an exploration module. \ac{TREx} allows one to easily swap out the sub-algorithms used for repetition and exploration.  In the first step, we model the repetition and exploration behavior separately to get candidates from both sources. Then, we generate the recommended basket from those candidates in the second step. 
The main architectural differences between previous approaches to the \ac{NBR} problem, which typically consists of a single treatment of all items, and \ac{TREx}, which treats repeat and explore items differently.
The pseudo-code for \ac{TREx} is given in Algorithm~\ref{alg:rfgr}. 
Next, we describe the three modules that make up \ac{TREx}.\footnote{Theoretically, \acs{TREx} allows us to choose or design the suitable repetition and exploration modules both targeted at the accuracy to achieve \acl{SOTA} performance. However, we aim to investigate the ``short-cut'' and relationship between accuracy and various beyond-accuracy metrics.}

\begin{algorithm}[!t]
	\SetAlgoLined
	\KwData{Basket sequence $S$, basket size $k$, repetition confidence threshold $v$}
	\KwResult{Recommended basket $B_u^{t+1}$ for each user $u$, }
	
	Calculate the repetition feature $\mathit{RepI}(i)$\ for each item\; \label{alg:line1}
	\For{each user $u$}{ \label{alg:line2}
		Get repeat items $I_{u,t}^\mathit{rep}$, and explore items $I_{u,t}^\mathit{expl}$\; \label{alg:line3}
		Calculate the repetition score $\mathit{RepS}^{u}(i)$ for each $i\in I_{u,t}^\mathit{rep}$\; \label{alg:line4}
		Remove items $i$ from $I_{u,t}^\mathit{rep}$, when $\mathit{RepS}^{u}(i) < v$\;  \label{alg:line5}
		
		Rank $I_{u,t}^\mathit{rep}$ according to $\mathit{RepS}^{u}(i)$ in descending order\; \label{alg:line6}
		Initialize next basket $B_u^{t+1}$\; \label{alg:line7}
		\eIf{$|I_{u,t}^\mathit{rep}|<k$}{ \label{alg:line8}
			Fill $B_u^{t+1}$ using $I_{u,t}^\mathit{rep}$\; \label{alg:line9}
			m $\leftarrow$ $k-|I_{u,t}^\mathit{rep}|$\; \label{alg:line10}
			Fill m empty slots of $B_u^{t+1}$ using explore items via exploration module\; \label{alg:line11}
		}{ \label{alg:line12}
			Fill $B_u^{t+1}$ using top-$k$ of $I_{u,t}^\mathit{rep}$\;\label{alg:line13}
		} \label{alg:line14}
	} \label{alg:line15}
	\caption{\OurMethod{} Framework }
	\label{alg:rfgr}
\end{algorithm}

\vspace*{-2mm}
\subsection{Repetition module}
As the repetition task is a much simpler task than exploration, we therefore design a repetition module targeted at improving the accuracy.
Intuitively, if a user consumed an item several times in the past, they are likely to repurchase that item in the next basket. 
Thus, frequency information is a strong signal for repetition prediction~\citep{triple2vec}. The \acf{PIF} introduced in TIFUKNN~\citep{tifuknn} and the recency window in UP-CF@r\citep{recency} both capture temporal dependencies by focusing more on recent behavior. 
However, they do not capture the item characteristics w.r.t.\ repurchasing. 
For example, a purchase of a bottle of milk and a pan is more likely to be followed by a repurchase of milk rather than a pan, even if both currently have the same purchase frequency. 
To consider both item features and user interest simultaneously, we use the repetition score $\mathit{RepS}^u(i)$ to represent the repurchase score of item $i$ for user $u$. This score is decomposed into two parts, the item-specific repurchase feature $\mathit{RepI}(i)$ and the user's interest $E_i^u$ in item $i$. Formally:
\begin{equation}
\mathit{RepS}^u(i) = E_i^u\cdot \mathit{RepI}(i)~.
\end{equation}
This corresponds to line~\ref{alg:line4} in Algorithm~\ref{alg:rfgr}.
 Given the items in the dataset $I = \{i_1, i_2, \ldots, i_m\}$, we need to derive the repurchase feature $\mathit{RepI}(i)$ for each item in the training set. 
First, the repurchase frequency $Rep^F(i)$ can be calculated by gathering the statistical information across users. To mitigate the impact of abnormally high values in some users, we introduce a hyperparameter $\alpha$ to discount the repurchase frequency of item $i$.
\begin{equation}
    \mathit{Rep}^F(i) = \frac{\sum_{U} \left(\text{item $i$ repurchase frequency}\right)^{\alpha}}{\# \text{users who bought item $i$ at least once}}~.
\end{equation}
In addition, some items might only have a few samples, which might lead to low confidence about their repetition feature estimation. We leverage the average estimate $\overline{\mathit{RepF}}$ across all items as supplementary information to help items with a few samples. Then, the final repetition feature is given by:
\begin{equation}
    \mathit{RepI}(i) = \mathit{Rep}^F(i) + \frac{\overline{\mathit{RepF}}}{N_i},
\end{equation}
where $N_{i}$ is the number of users who bought item $i$. Thus, the average $\overline{\mathit{RepF}}$ will have a small effect on $\mathit{RepI}(i)$ when we have more samples to compute item-specific features. This corresponds to line~\ref{alg:line1} in Algorithm~\ref{alg:rfgr}.

The item frequency in a user's historical baskets can partially reflect the user's interest. 
Yet, user interests can also be dynamic. 
To model temporal dependencies, we introduce a time-decay factor $\beta$, which makes the recent interactions have more impact on the interest $E_i^u$. Assume that a specific item $i$ was purchased by the user $u$ several times in their historical baskets $\{B_u^{l_1}, B_u^{l_2}, \ldots, B_u^{l_m}\}$; the corresponding position set is denoted as $L_i = \{l_1, l_2, \ldots, l_m\}$; then $E_i^u$ is defined as:
\begin{equation}
\textstyle
   E_i^u = \sum_{j=1}^m \beta^{T-l_j}~,
\end{equation}
where $T$ represents the length of the user's basket. 
\ac{TREx}'s repeat recommendation model takes item features, user interests, and the temporal order of baskets into consideration.
We treat the items in baskets independently and calculate the repetition score $\mathit{RepS}$ for all items that appeared in the previous baskets for each user, which will be used in the final basket generation process.

\vspace*{-2mm}
\subsection{Exploration module}

As it is more challenging than repetition, exploration is also an important aspect of \ac{NBR}. 
To complement the repetition module, we design different exploration modules, targeting item fairness and diversity, respectively. 
For each user $u$, the exploration candidates $I_{u, t}^{expl}$ are the set of items that the user never bought before. 

\header{Item fairness} According to~\citep{nbr-rep-expl}, we find that \ac{NBR} methods usually have varying degrees of popularity bias, which means they recommend more popular items compared to the ground truth and harm item fairness. Thus, we recommend unpopular items $i\in G^{-}$for the exploration module for the sake of approaching the distribution of ground truth and decreasing the exposure gap between the popular and the unpopular groups. Specifically, we randomly sample explore items based on a sampling probability, which is calculated from the purchase frequency of unpopular items.

\header{Diversity} Diversity optimizes for more dispersed categories in the predicted basket. For each user, we record categories of repetition candidates, rank exploration candidates according to their popularity, and select explore items to fill in the $B_u^{t+1}$ in turn. The category of each explore item differs from the categories already in $B_u^{t+1}$.

\vspace*{-2mm}
\subsection{Basket generation module}
To construct the final basket to be recommended by \ac{TREx} for the accuracy objective, we adopt a repetition greedy approach and first consider the item candidates generated by the repetition module and fill the remaining slots via the exploration module. $\mathit{\acs{TREx}_{Fairness}}$ and $\mathit{\acs{TREx}_{diversity}}$ denote \acs{TREx} with the exploration module targeted at fairness and diversity, respectively.
For a user $u$, we get their repetition score $\mathit{RepS^{u}}(i)$, where $i\in I_{u,t}^\mathit{rep}$ (Algorithm~\ref{alg:rfgr}, lines~\ref{alg:line3}--\ref{alg:line4}). 
First, we define a confidence threshold $v$ for the repetition score and repetition items are removed from the $i\in I_{u,t}^\mathit{rep}$ when the corresponding $\mathit{RepS^{u}}(i) < v$ (line~\ref{alg:line5}).\footnote{The confidence threshold $v$ controls the proportion of repeat items and explore items in the recommendation, as well as the accuracy and beyond-accuracy trade-off in this paper. We sweep repetition confidence bound $v$ to get \acs{TREx} variants with different accuracy and beyond-accuracy metrics performance.}
Then, $I_{u,t}^\mathit{rep}$ can be seen as the repetition candidates set.
If the number of repetition candidates exceeds the basket size, the items with a high score will have priority to fill the basket (Algorithm~\ref{alg:rfgr}, line~\ref{alg:line13}). 
If the number of repetition candidates is smaller than the basket size, the basket is first filled with all items in the repetition candidates set $I_{u,t}^\mathit{rep}$. Then, we fill up the basket using the explore items via the exploration module, where $m$ represents the number of empty slots (lines~\ref{alg:line9}--\ref{alg:line11}).

\vspace*{-2mm}
\section{Experiments}

\begin{table}
\newcommand{\tabincell}[2]{\begin{tabular}{@{}#1@{}}#2\end{tabular}}
  \caption{Statistics of the processed datasets. }
  \label{dataset}
  \setlength{\tabcolsep}{2pt}
  \begin{tabular}{l@{}cccccc}
    \toprule
    \bf Dataset & \bf \#items & \bf \#users & \tabincell{c}{\bf Avg. \\ \bf basket\\\bf size} & \tabincell{c}{\bf Avg. \\ \bf \#baskets\\\bf per user}  & \tabincell{c}{\bf Repeat\\\bf ratio} & \tabincell{c}{\bf Explore\\\bf ratio}\\
    \midrule
    Instacart & 29,399 & 19,210 & 10.06 & 15.91 & 0.60 & 0.40\\
    Dunnhumby & 37,162 & \phantom{0}2,482 & 10.07 & 43.17 & 0.43 & 0.57\\
    \bottomrule
  \end{tabular}
\end{table}

\subsection{Experimental setup}
\begin{figure}[h]
\centering
    \includegraphics[width=1\linewidth]{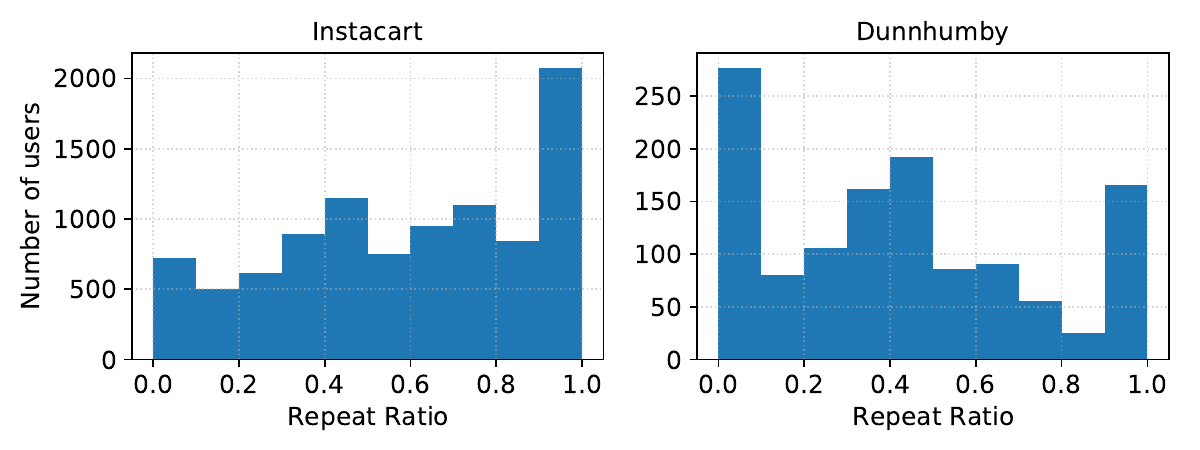}
    \caption{Distribution of users across different repeat ratios for Instacart and Dunnhumby.}
    \label{fig:user_dist_repeat_ratio}
\end{figure}

\textbf{Datasets.}
We conduct experiments on two widely-used datasets:
\begin{enumerate*}[label=(\roman*)]
\item Instacart,\footnote{\url{https://www.kaggle.com/c/instacart-market-basket-analysis/data}} which includes a large number of grocery orders from users; following~\citep{liu-2024-measuring, naumov2023time}, ${\sim}$20000 users are randomly selected to conduct experiments; and
\item Dunnhumby,\footnote{\url{https://www.dunnhumby.com/source-files/}} which contains two years' household-level transactions of 2500 frequent shoppers at a retailer.
\end{enumerate*}
Following \cite{liu-2024-measuring, recanet}, we sample users who have at least three baskets and remove items that appeared less than five times. 
The two datasets vary in the repeat ratio, i.e., the proportion of repeat items in the ground-truth baskets~\citep{nbr-rep-expl}.
We focus on the fixed size (10 or 20) NBR problem. The statistics of the processed datasets are shown in Table \ref{dataset}.
In our experiments, each dataset is partitioned according to~\citep{naumov2023time, recanet, recency, liu-2024-measuring}. 
The training baskets encompass all user baskets except the last one. In cases where users have over 50 baskets in the training data, only their last 50 baskets are considered for inclusion in the training set. The final baskets of all users are then divided equally between a 50\% validation set and a 50\% test set. Figure~\ref{fig:user_dist_repeat_ratio} shows the distribution of users across repeat ratios, which is the proportion of \emph{repeat items} in the ground-truth basket.

\header{\acs{NBR} baselines}
We compare \acs{TREx} with 8 representative baselines, which we select based on their characteristics in the analysis performed in~\cite{nbr-rep-expl, liu-2024-measuring}, divided into three groups:

\vspace*{-1mm}
\subsubsection{Simple baselines}
\begin{enumerate*}[label=(\roman*)]
    \item \textbf{G-TopFreq} uses the $k$ most popular items in the dataset to form the recommended next basket. 
    \item \textbf{P-TopFreq} is a personalized TopFreq method, which treats the most frequent $k$ items in historical records of the user as the next basket. %
    \item \textbf{GP-TopFreq}~\citep{nbr-rep-expl} is a simple combination of P-TopFreq and G-TopFreq, which first use P-TopFreq to fill the basket, then use G-TopFeq to fill the remaining slots.
\end{enumerate*}

\vspace*{-1mm}
\subsubsection{Nearest neighbor-based methods}
\begin{enumerate*}[label=(\roman*)]
    \item \textbf{TIFUKNN}~\citep{tifuknn} is a state-of-art method that models the temporal dynamics of frequency information of users' past baskets to introduce Personalized Frequency Information (PIF), then it uses KNN-based method on the PIF.
    \item \textbf{UP-CF@r}~\citep{recency} is a combination of recency aware user-wise popularity and user-wise collaborative filtering.
\end{enumerate*}

\vspace*{-1mm}
\subsubsection{Neural network-based methods}
\begin{enumerate*}[label=(\roman*)]
  \item \textbf{Dream}~\citep{dream} models users' global sequential basket behavior for \acs{NBR} using \ac{RNN}. 
  \item \textbf{DNNTSP}~\citep{dnntsp} is a state-of-art method that leverages a GNN and self-attention techniques. It encodes item-item relations via a graph and employs a self-attention mechanism to capture temporal dependencies of users' basket sequences.
  \item \textbf{ReCANet}~\citep{recanet} is a repeat-only model for NBR, which uses user-item representations with historical consumption patterns via \acs{RNN}.
\end{enumerate*}

\header{Configurations}
To assess group fairness (Section~\ref{sec: metric}), we follow configurations from previous research~\citep{li2022fairness, liu-2024-measuring}; the group of items is determined by their popularity (i.e., the number of purchases recorded in the historical baskets of the dataset). The top 20\% of items with the highest purchase frequency as the popular group ($G^+$), while the remaining 80\% of items are assigned to the unpopular group ($G^-$).
For the baseline methods, a grid search is performed to find the optimal hyper-parameters via the validation set.
For TIFUKNN, the number of neighbors $k$ is tuned on $\{100, 300, 500, 900, 1100, 1300\}$, the number of groups $m$ is tuned on $\{3, 7, 11, 15, 19, 23\}$, the within-basket time-decayed ratio $r_b$ and the group time-decayed ratio $r_g$ are selected from $\{0.1, 0.2, \ldots, 0.9, 1\}$, and the fusion weight $\alpha$ is selected from $\{0, 0.1, \ldots, 0.9, 1\}$. For UP-CF@r, recency window $r$ is tuned on $\{1, 5, 10, 25, 100, \infty\}$, locality $q$ is tuned on $[1, 5, 10, 50, 100, \allowbreak 1000]$, and asymmetry $\alpha$ is tuned on $\{0, 0.25, 0.5, 0.75, 1\}$. For Dream, DNNTSP, and ReCANet, the item and user embedding size is tuned on $\{16,32,64,128\}$.
As to \ac{TREx}, for the repetition module, $\alpha$ is selected from \{0, 0.1, 0.2, 0.3, 0.4, 0.5, 0.6, 0.7, 0.8, 0.9, 1.0\}, and the time-decay factor $\beta$ is selected from \{0.7, 0.75, 0.8, 0.85, 0.9, 0.95, 1.0\}. To facilitate reproducibility, we release the source code and all hyper-parameters in an online repository:  \url{https://github.com/lynEcho/TREX}.

\begin{table*}
\centering

\caption{Comparison of TREx-Rep (repetition-module only) against baselines and two types of state-of-art methods; boldface indicates the maximum; underlining indicates the second best performing method. $\dagger$ indicates that TREx-Rep results achieve the same level of performance as \acs{SOTA} baselines (paired t-test).}
\label{tab:performance-acc}
\begin{tabular}{@{} l l c c c c c c c c c}
\toprule
Dataset& Metric& G-TopFreq& P-TopFreq&GP-TopFreq&UP-CF@r& TIFUKNN &Dream &DNNTSP&ReCANet&TREx-Rep\\
\midrule
\multirow{6}{*}{\rotatebox[origin=c]{90}{Instacart}}& Recall@10 &0.0704 &0.3143 &0.3150 &0.3377 &0.3456 &0.0704 &0.3295 &\textbf{0.3490} &\underline{0.3476}$\dagger$\\
& NDCG@10 &0.0817 &0.3339 &0.3343 &0.3582 &0.3657 &0.0817 &0.3434 &\textbf{0.3699} &\underline{0.3661}$\dagger$\\
& PHR@10 &0.4600 &0.8447 &0.8460 &0.8586 &0.8639 &0.4600 &0.8581 &\textbf{0.8668} &\underline{0.8655}$\dagger$\\
\cmidrule{2-11}
& Recall@20 &0.0973 &0.4138 &0.4168 &0.4405 &\underline{0.4559} &0.0979 &0.4339 &\textbf{0.4562} &0.4557$\dagger$\\
& NDCG@20 &0.0962 &0.3889 &0.3902 &0.4161 &\underline{0.4271} &0.0968 &0.4018 &\textbf{0.4303} &0.4269$\dagger$\\
& PHR@20 &0.5302 &0.8921 &0.8959 &0.9045 &\textbf{0.9098} &0.5346 &0.9033 &\underline{0.9097} &0.9092$\dagger$\\
\midrule
\multirow{6}{*}{\rotatebox[origin=c]{90}{Dunnhumby}}& Recall@10 &0.0897 &0.1628 &0.1628 &0.1699 &\underline{0.1763} &0.0896 &0.0871 &0.1730 &\textbf{0.1815}$\dagger$\\
& NDCG@10 &0.0798 &0.1562 &0.1562 &0.1639 &\underline{0.1683} &0.0759 &0.0792 &0.1625 &\textbf{0.1689}$\dagger$\\
& PHR@10 &0.3795 &0.5399 &0.5399 &0.5536 &\underline{0.5729} &0.3873 &0.4303 &0.5655 &\textbf{0.5761}$\dagger$\\
\cmidrule{2-11}
& Recall@20 &0.1046 &0.2075 &0.2075 &0.2168 &0.2227 &0.1081 &0.1442 &\underline{0.2252} &\textbf{0.2257}$\dagger$\\
& NDCG@20 &0.0877 &0.1787 &0.1787 &0.1885 &\underline{0.1917} &0.0853 &0.1021 &0.1879 &\textbf{0.1921}$\dagger$\\
& PHR@20 &0.4392 &0.6116 &0.6116 &0.6326 &0.6342 &0.4558 &0.5378 &\underline{0.6377} &\textbf{0.6390}$\dagger$\\
\bottomrule
\end{tabular}
\end{table*}

\begin{figure}[t]
    \centering
    \includegraphics[width=\linewidth]{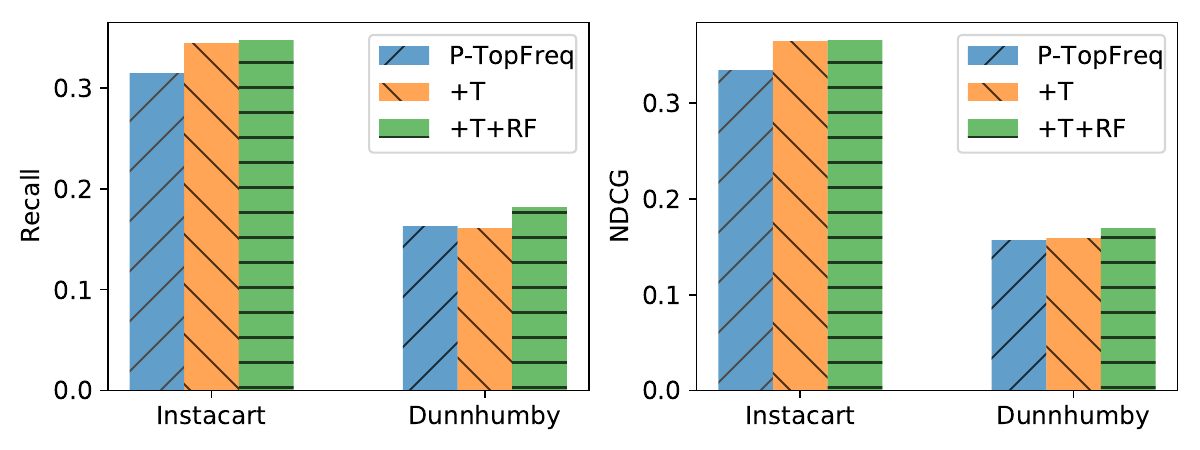}
    \caption{Performance of TREx-Rep when we add a time-decay factor $\beta$ (+T), add both $\beta$ and item-specific repetition feature $RepI(i)$ (+T+RF).}
    \label{fig:rep-perf}
\end{figure}

\subsection{Overall accuracy performance}

By decoupling the repetition and exploration tasks, \ac{TREx}-Rep optimizes for the repeat items prediction and accounts for the accuracy of the NBR performance.
Table~\ref{tab:performance-acc} shows the experimental results for \ac{TREx}-Rep and the baselines.
We observe that \ac{TREx}-Rep surpasses two complex deep learning-based methods (i.e., Dream and DNNTSP) by a large margin on the Dunnhumby and Instacart datasets, and \ac{TREx}-Rep always achieves or matches the \acs{SOTA} accuracy on both datasets across different accuracy metrics.
Note that, \ac{TREx}-Rep achieves a competitive accuracy performance by only using part of the available slots in the basket.\footnote{As TREx-Rep only recommends repeat items, the basket could not be fulfilled when the number of user's repeat items (historical items) is smaller than the basket size. ReCANet also only recommends repeat items, however, it is a complex neural-based model, which is much slower than the proposed TREx-Rep module.}
Compared to the deep learning methods with complex architectures that try to learn basket representations and model temporal relations, TREx-Rep is very efficient due to its simplicity.

\begin{figure}[t]
    \centering
    \includegraphics[width=0.5\linewidth]{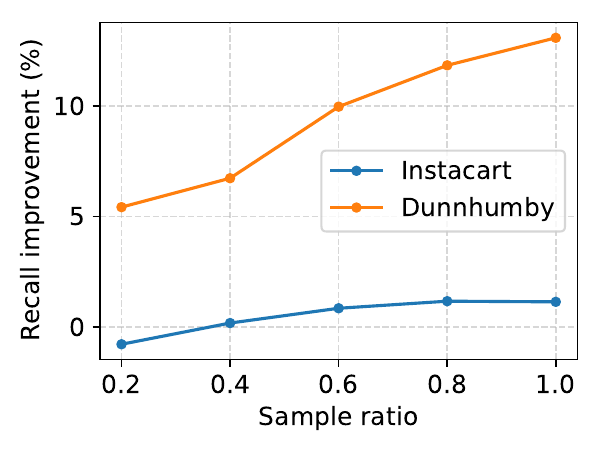}
    \caption{The recall improvement of (+T+RF) over (+T) when the training sample ratio changes from 0.2 to 1.}
    \label{fig:rep-repfeature}
\end{figure}

To investigate the effect of the repetition features and the improvement in repetition performance in \ac{NBR}.
We conduct experiments on TREx-Rep by gradually adding the time-decay factor $\beta$ and item-specific repetition feature $\mathit{RepI}(i)$. The results are shown in Figure~\ref{fig:rep-perf}. The accuracy increases when we gradually integrate different factors into \ac{TREx}-Rep, which indicates that both the time-decay factor $\beta$ and the item-specific repetition feature $RepI(i)$ contribute to the accuracy performance of \ac{TREx}-Rep.
Significant improvements over only using the time-decay factor $\beta$ can be observed on the Dunnhumby dataset when the item-specific repetition feature $RepI(i)$ is also adopted to compute the repetition score $RepS^u(i)$. 
Note that the improvement of adding $RepI(i)$ to \ac{TREx}-Rep on the Instacart dataset is relatively small. We conjecture that items in the Instacart dataset are more regular products, that have little difference in repetition feature with each other. 
Figure~\ref{fig:rep-repfeature} shows the performance when using different amounts of training samples, the improvement in recall resulting from adding $RepI(i)$ increases when we use more training data since we have more samples for estimating the repetition feature $RepI(i)$.

\subsection{Beyond-accuracy performance}
We conduct experiments to verify whether \acs{TREx} with the designed models (i.e., $\acs{TREx}_{Diversity}$ and $\acs{TREx}_{Fairness}$) could achieve better performance on representative diversity and item fairness metrics.
Note that, the recommended basket remains fixed for a specific user in existing baselines, resulting in fixed performance regarding both accuracy and beyond-accuracy metrics on each dataset.
In contrast, \acs{TREx} provides the flexibility to adjust the trade-off between accuracy and beyond-accuracy metrics by adjusting the repetition confidence bound $v$. This allows for a more nuanced control over the recommendation process compared to traditional baselines.

\begin{figure}
    \centering
    \includegraphics[clip,trim=55mm 175mm 55mm 0mm,width=\columnwidth]{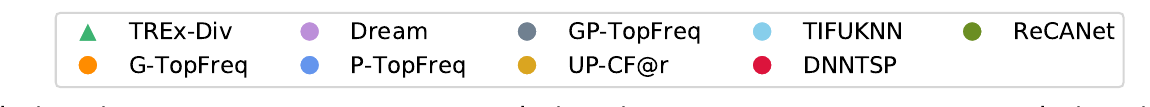}
    \\
    \includegraphics[clip,trim=11mm 0mm 25mm 18mm,width=\columnwidth]{figures/diversity.pdf}
    \caption{Performance of $\acs{TREx}_{Diversity}$ at different $v$ values, compared with different \acs{NBR} methods in terms of different diversity metrics. The red $+$ marker indicates the direction with both high accuracy and diversity.}
    \label{fig:diversity-perf}
\end{figure}

\begin{figure*}
    \centering
    \includegraphics[width=0.85\linewidth]{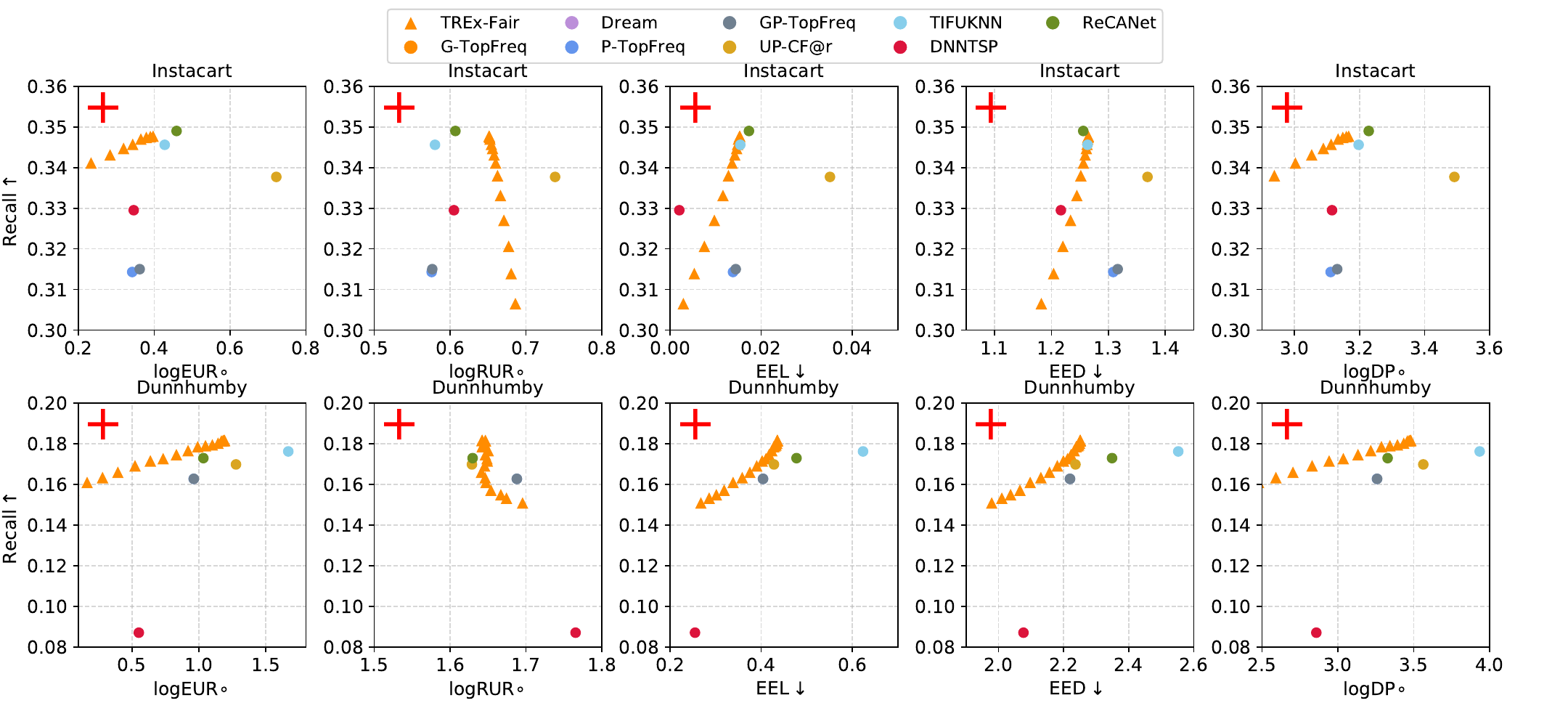}
    \vspace*{-1mm}
    \caption{Performance of $\acs{TREx}_{Fairness}$ at different $v$ values, compared with different \acs{NBR} methods in terms of different fairness metrics. The red $+$ marker indicates the direction with both high accuracy and fairness.}
    \label{fig:fairness-perf}
\end{figure*}

\header{Diversity}
The experimental results w.r.t.\ the accuracy and different diversity metrics (i.e., ILD, Entropy, and DS) are shown in Figure~\ref{fig:diversity-perf}.\footnote{G-TopFreq and Dream exhibit low recall, fairness, and diversity, which prevents them from being visible in Figures~\ref{fig:diversity-perf} and \ref{fig:fairness-perf}.} We have the following observations:
\begin{enumerate*}
    \item Compared to methods (i.e., TIFUKNN and ReCANet) with the best accuracy, $\mathit{\acs{TREx}_{Diversity}}$ can achieve better performance in terms of all three diversity metrics while preserving the same level of accuracy on both datasets.
    \item In contrast to other baseline methods (excluding TIFUKNN and ReCANet), $\mathit{\acs{TREx}_{Diversity}}$ showcases the ability to recommend baskets with enhanced accuracy and diversity simultaneously.
\end{enumerate*}

\header{Item fairness}
The experimental results regarding the accuracy and five fairness metrics (LogRUR, logEUR, logDP, EEL, and EED) are depicted in Figure~\ref{fig:fairness-perf}. 
Based on our analysis, we make the following observations:
\begin{enumerate*}[label=(\roman*)]
    \item On the Dunnhumby dataset, $\acs{TREx}_{Fairness}$ demonstrates superior fairness w.r.t.\ logDP and logEUR while maintaining the same level of accuracy performance as the best-performing baselines (i.e., TIFUKNN and ReCANet). Similarly, on Dunnhumby, $\acs{TREx}_{Fairness}$ showcases enhanced fairness across four fairness metrics (logDP, logEUR, EEL, and EED) while achieving accuracy performance comparable to the best-performing baselines.
    \item $\mathit{\acs{TREx}_{Fairness}}$ demonstrates its capability to recommend baskets with improved accuracy and fairness w.r.t.\ logDP and logEUR concurrently, when compared to complex baselines such as Dream, UP-CF@r, and DNNTSP.
    \item In terms of logRUR, $\mathit{\acs{TREx}_{Fairness}}$ exhibits inferior performance in fairness while maintaining similar accuracy levels compared to several existing baselines. Moreover, as both accuracy and fairness decrease simultaneously, a win-win and lose-lose scenario is evident rather than a conventional trade-off relationship in this fairness evaluation.
\end{enumerate*}

\header{Connections with accuracy}
To get a better understanding of the possibility of leveraging the ``short-cut'' via TREx to improve beyond-accuracy metrics, we conduct an analysis by categorizing these beyond-accuracy metrics into different groups based on their connections with accuracy (see Section~\ref{sec: metric} and Table~\ref{metric_table}).

We can observe that $\mathit{\acs{TREx}}$ can easily achieve better performance w.r.t.\ beyond-accuracy metrics have no connections with the accuracy (i.e., ILD, Entropy, DS, and logDP) on two datasets. When beyond-accuracy metrics (e.g., logEUR, EEL, and EED) exhibit weak associations with accuracy, $\mathit{\acs{TREx}}$ outperforms alternative methods in some instances (4 out of 6). However, in cases where beyond-accuracy metrics are strongly correlated with accuracy (e.g., logRUR), $\mathit{\acs{TREx}}$ struggles to achieve superior performance. Since only accurate predictions contribute to improvements in logRUR fairness, leveraging the exploration module to optimize such beyond-accuracy metrics is very challenging.

\vspace*{-2mm}
\subsection{Reflections and discussions}

The above results verify our hypothesis and demonstrate the effectiveness of leveraging a ``short-cut'' strategy to achieve better beyond-accuracy under the current evaluation paradigms. 

It is controversial to use this ``short-cut'' strategy in real-world scenarios when \acs{NBR} practitioners consider beyond-accuracy metrics.
In scenarios where the accuracy of exploration is not important to practitioners and only overall accuracy is of concern, the ``short-cut'' strategy proves to be a straightforward and efficient means to achieve better performance w.r.t.\ various beyond-accuracy metrics. 
\acs{TREx} must be considered or serve as a baseline before designing more sophisticated methods, such as including multi-objective loss functions~\citep{nbrdiversity, chen2020improving}, integer programming~\citep{zhao2023fairness}, and so on.

However, in some scenarios, it is unreasonable to sacrifice the exploration accuracy despite it being low. Therefore, the existence of the ``short-cut'' strategy reveals the potential flaws of the existing evaluation paradigms (i.e., using overall metrics to define success). We look into the exploration accuracy~\citep{nbr-rep-expl} of $\mathit{\acs{TREx}_{Diversity}}$ when it outperforms several existing baselines in terms of both overall accuracy and diversity (i.e., success according to existing evaluation paradigm). 
Table~\ref{tab:expl_acc} shows the huge decrease in the accuracy of exploring items in the recommended basket of $\mathit{\acs{TREx}_{Diversity}}$, compared to these baselines, since the designed module in $\mathit{\acs{TREx}_{Diversity}}$ is mainly designed for improving diversity and does not consider accuracy.
In this sense, we can not simply claim the superiority of $\mathit{\acs{TREx}_{Diversity}}$ compared to these baselines just depends on the overall performance.

Note that, the fundamental reason for the existence of this ``short-cut'' is that predicting accurate \emph{explore items} is much more difficult than predicting \emph{repeat items}, and exploration prediction only accounts for a limited user's overall accuracy~\citep{nbr-rep-expl, li-2023-repetition, li-2023-repetition-offline, li-2023-will}. \emph{Given that exploration prediction contributes only minimally to the overall accuracy of users, it becomes feasible to allocate resources toward optimizing other beyond-accuracy metrics instead of accuracy itself.}

Therefore, beyond using the overall performance to measure accuracy and beyond-accuracy metrics, a fine-grained level evaluation could help to provide a more rigid identification of the success when considering beyond-accuracy metrics.

\begin{table}[h]
\centering
\caption{Exploration accuracy~\cite{nbr-rep-expl} of $\acs{TREx}_{Diversity}$ compared with \acs{NBR} methods that are inferior to it within existing evaluation paradigms.}
\label{tab:expl_acc}
\small
\setlength{\tabcolsep}{1.5mm}
\begin{tabular}{@{}clccccc@{}}
\toprule
Dataset & Metric & TIFUKNN & Dream & DNNTSP  & TREx-Div \\  \midrule
\multirow{4}{*}{\rotatebox[origin=c]{90}{Instacart}} &  $\mathrm{Recall}_{expl}@10$  & 0.0014 & 0.0322 & 0.0014  & 0.0002 \\
                  & $\mathrm{PHR}_{expl}@10$       & 0.0037 & 0.1431 & 0.0040 & 0.0009  \\ \cmidrule(l){2-6} 
                  & $\mathrm{Recall}_{expl}@20$        & 0.0077 & 0.0526 & 0.0072  & 0.0008 \\
                  &  $\mathrm{PHR}_{expl}@20$      & 0.0198 & 0.2120   & 0.0217 & 0.0031 \\  \midrule
\multirow{4}{*}{\rotatebox[origin=c]{90}{Dunnhumby}} & $\mathrm{Recall}_{expl}@10$       & 0.0042  & 0.0111 & 0.0017  & 0.0000   \\
                  &  $\mathrm{PHR}_{expl}@10$       & 0.0139  & 0.0521 & 0.0085  & 0.0019 \\  \cmidrule(l){2-6} 
                  &  $\mathrm{Recall}_{expl}@20$      & 0.0069 &0.0214  & 0.0028  & 0.0016 \\
                  & $\mathrm{PHR}_{expl}@20$      & 0.0232 & 0.1045 & 0.0115  & 0.0065  \\  \midrule
\end{tabular}
\end{table}

\vspace*{-2mm}
\section{Conclusion}
We have expanded the research objectives of \acs{NBR} to go beyond sole accuracy to encompass both accuracy and beyond-accuracy metrics.
We have recognized a potential ``short-cut'' strategy to optimize beyond-accuracy metrics while preserving high accuracy levels.
To capitalize on and validate the presence of such ``short-cuts,'' we have introduced a plug-and-play framework called \acfi{TREx} considering the differences between repetition and exploration tasks. This framework treats repeat items and explore items as distinct entities, employing a straightforward yet highly effective repetition module to uphold accuracy standards. Concurrently, two exploration modules have been devised to target the optimization of beyond-accuracy metrics.
We have conducted experiments on two publicly available datasets w.r.t.\ eight representative beyond-accuracy metrics, including item fairness (i.e., logEUR, LogRUR, logDP, EEL, and EED) and diversity (i.e., ILD, Entropy, and DS). 

Our experimental results demonstrate the effectiveness of our proposed ``short-cut'' strategy, which can achieve better beyond-accuracy performance w.r.t.\ several fairness and diversity metrics on different datasets. 
Additionally, we group beyond-accuracy metrics according to the strength of their connection with accuracy. 
Our analysis reveals that the stronger the connection with accuracy, the more difficult it becomes to employ a ``short-cut'' strategy to optimize these beyond-accuracy metrics, favoring the metrics with a stronger connection to avoid such short-cuts.

As to the broader implications of our work, we have discussed the reasonableness of leveraging the ``short-cut'' strategy to trade the accuracy of exploration for beyond-accuracy metrics in various scenarios. 
The presence of this ``short-cut'' highlights a potential flaw in the definition of success within existing evaluation paradigms, particularly in scenarios where exploration accuracy is important despite being low~\citep{williams-2014-emotions}. A fine-grained level evaluation should be performed in \acs{NBR} to offer a more precise identification of achieving ``better'' performance in such a scenario.

Despite the simplicity of the ``short-cut'' strategy and \acs{TREx}, our paper sheds light on the research direction of considering both accuracy and beyond-accuracy metrics in \acs{NBR}. Rather than blindly embracing sophisticated methods in \acs{NBR}, follow-up research should realize the existence of the ``short-cut'' and potential flaws of existing evaluation paradigms in this research direction.

\subsubsection*{\bf Acknowledgements}
This work is partially supported by the Dutch Research Council (NWO), under project numbers 024.004.022, NWA.\-1389.20.183, KICH3.LTP.20.006, and VI.Vidi.223.166.
All content represents the opinion of the authors, which is not necessarily shared or endorsed by their respective employers and/or sponsors.

\clearpage

\bibliographystyle{ACM-Reference-Format}
\bibliography{references}

\end{document}